\documentclass{article}
\usepackage[utf8]{inputenc}

\usepackage[a4paper,margin=3cm,top=2.5cm]{geometry}
\usepackage{titling}
\usepackage[T1]{fontenc}
\usepackage{libertinus}
\usepackage{microtype}
\usepackage{csquotes}
\usepackage[all]{nowidow}

\usepackage[inline]{enumitem}
\setlist{nolistsep}
\newlist{inlinelist}{enumerate*}{1}
\setlist[inlinelist,1]{label=\textit{\roman*)}}
\newlist{inlineabc}{enumerate*}{1}
\setlist[inlineabc,1]{label=\textit{\alph*)}}

\usepackage{graphicx}
\usepackage[font=small]{caption}
\usepackage{booktabs}
\usepackage[hyphens]{url}
\usepackage[hidelinks]{hyperref}

\usepackage[capitalize,noabbrev,nameinlink]{cleveref}
\usepackage[sortcites]{biblatex}
\bibliography{references}

\renewenvironment{abstract}{\small\quotation}{\endquotation}

\usepackage[dvipsnames,svgnames]{xcolor} 
\usepackage[normalem]{ulem} 

\makeatletter
\font\uwavefont=lasyb10 scaled 700
\def\spelling{\bgroup\markoverwith{\lower3.5\p@\hbox{\uwavefont\textcolor{Red}{\char58}}}\ULon}
\def\grammar{\bgroup\markoverwith{\lower3.5\p@\hbox{\uwavefont\textcolor{LimeGreen}{\char58}}}\ULon}
\def\phrasing{\bgroup\markoverwith{\lower3.5\p@\hbox{\uwavefont\textcolor{RoyalBlue}{\char58}}}\ULon}

\newcommand\remove{\bgroup\markoverwith{\textcolor{red}{\rule[0.5ex]{2pt}{0.4pt}}}\ULon}
\newcommand\insertion{\bgroup\markoverwith{\textcolor{Green}{\rule[-0.5ex]{2pt}{0.6pt}}}\ULon}
\makeatother

\title{
  \LARGE Lifting the Cage of Consent\\
  \large A~Techno-Legal Perspective on Evolvable Trust Relationships
  \vspace{-.5\baselineskip}
}
\author{Beatriz Esteves \& Ruben Verborgh\\[-.25em]
  \small
  IDLab,
  Department of Electronics and Information Systems,
  Ghent University,
  Belgium}
\date{\vspace{-2\baselineskip}}

\begin{document}

\maketitle

\begin{abstract}
\noindent
Those concerned about privacy worry that
personal data changes hands too easily.
We argue that the actual challenge is the exact opposite:
our data does not flow well enough,
cultivating a~reliance on questionable and often unlawful shortcuts
in a~desperate bid to survive within today's data-driven economy.
Exclusively punitive interpretations of
protective legislation such as the GDPR
throw out the baby with the bathwater
through barriers that equally hinder
\enquote{doing the right~thing}
and
\enquote{doing the wrong~thing},
\linebreak
in an abject mistranslation of how ethical choices
correspond to financial cost.
As~long as privacy-friendly data treatment
proves more expensive or complicated
than readily available alternatives,
economic imperatives will continue to outrank their legal counterparts.
We examined existing legislation
with the aim of facilitating mutually beneficial interactions,
rather than more narrowly focusing on the prevention of undesired behaviors.
In~this article,
we propose the implementation of evolvable trust systems
as a~scalable alternative
to the omnipresent yet deeply broken delusion of ill-informed consent.
We~describe personalized, technology-assisted legal processes
for initiating and maintaining long-term trust relationships,
which enable parties to reliably and sustainably
exchange data, goods, and services.
Our proposal encourages a~redirection of additional efforts towards
the techno-legal alignment of economical incentives with societal~ones,
reminding us that---%
while trust remains an inherently human~concept---%
technology can support people
in evolving and scaling their relationships
to meet the increasingly complex demands
of current and future data~landscapes.
\end{abstract}

\section{The Cage of Consent}
\label{sec:Introduction}

\subsection{Privacy is a~red herring}
News media are eager to report on the next scandal
leaking from the industrial data kitchens
of Google, Adobe, ByteDance, and the like.
The fourth power's crusade to raise awareness of
the societal devastation inflicted by Cambridge Analytica~\cite{Hinds2020}
and subsequent disasters~\cite{cimpanu_database_2019,noauthor_ftc_2023,perappadan_us_2023,halliday_data_2024}
echoes across the economical and political wasteland
that sprouted such widespread abuse in the first~place.
In an embarrassing bout of~irony,
journalism's oversimplification of \emph{privacy}
as a~dial that must be cranked all the way~up,
inadvertently plays into the cards
of those same data~funneling companies
they believe to be~lecturing.
After Zuckerberg unilaterally proclaimed
the end of privacy in~2010~\cite{ZuckerbergGuardian2010},
Big Tech's more recent public communications
now smother our collective indignation
with an ostentatious appropriation of privacy~\cite{
  FacebookPrivacyGuardian2018,
  FacebookEncryptionGuardian2023,
  GooglePrivacyGuardian2021,
  MicrosoftPrivacy2014,
}
as if suddenly owed gratitude for their artful reinvention.
Such blatant \mbox{red herrings}
reinvigorate society's futile attempts at reclaiming privacy,
thereby spawning a~fatal terminological stretch
from which any leftover meaning bleeds in~vain.
Meanwhile,
companies project their own deeply rooted fear of data~starvation
onto people's genuine worry of exploitation---%
circularly fueled by a~compulsion
to manifest as merciless data~scavengers
at the faintest smell of digital~oil.

\begin{figure}
  \centering
  \includegraphics[width=\linewidth]{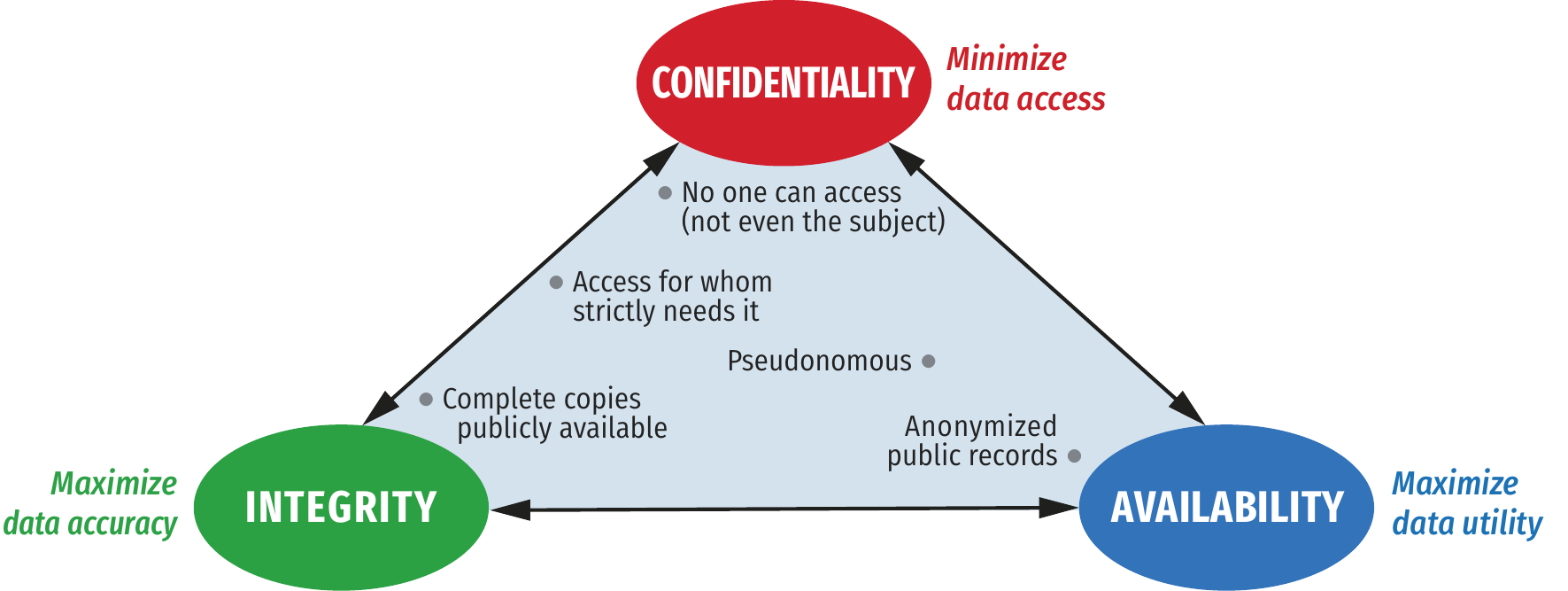}
  \caption{
    The CIA~triad~\cite{samonas2014cia} of
    \emph{Confidentiality}, \emph{Integrity}, and \emph{Availability}
    illustrates that personal data requirements
    involve nuance and complexity
    beyond what any one-dimensional \emph{privacy}~lever can model.
    One can never maximize all three;
    the appropriate degree of compromise
    is a~case-specific choice.
  }
  \label{fig:CIA-triad}
\end{figure}

We confidently discredit these companies' abundant mentions of privacy
as nothing but premeditated distractions,
because achieving complete privacy is a~straightforward goal
that carries surprisingly little utility for the majority of humanity.
Print all of your documents,
photos, data, medical files,
and place them together with your electronic devices
inside a~large mold.
Pour in an ample volume of concrete
before sinking the receptacle in the ocean,
and you will be a~few dozen GDPR erasure requests away
from no~one accessing your data ever~again.
Regrettably, that includes~yourself.
This contrived example highlights
the danger that arises when privacy-focused debates
ignore individuals' vastly more complex protection needs
beyond a~theoretical notion of isolation.
The \emph{CIA~triad}~\cite{samonas2014cia} in \cref{fig:CIA-triad}
more accurately expresses real-world data decisions
as a~continuous tension between
\emph{Confidentiality}, \emph{Integrity}, and \emph{Availability}.

What we often loosely refer to as \enquote{privacy}
lets itself be more precisely defined as \enquote{confidentiality},
a~preference for restricting the number of parties
who are privy to certain data.
Confidentiality unfortunately clashes with availability because,
while redundancy and backups improve the reliability of data~access,
they simultaneously leave more pathways for eavesdroppers.
Essentially,
whatever makes it hard or easy for others to retrieve our data,
affects ourselves in similar ways.
The same door~lock that delays seasoned burglars from entering,
also blocks rightful inhabitants who lost or forgot their house~key.

We must therefore carefully reflect on
\emph{when} and \emph{how} personal data contributes
to the realization of individuals' own objectives,
when aiming to construct a~legislative framework for
supporting people in their daily lives and with their long-term ambitions.
We cannot mistakenly assume that
100\%~confidentiality represents the universal optimum
within all data exchange relationships,
and must understand the other parts of the scale
in order to provide meaningful legal and technological empowerment.
Absence of such consideration
recasts progress as sophisticated damage control,
rather than advancing
the more fundamental and sustainable objective
of engaging in mutually beneficial transactions.

\subsection{Personal data unlocks mutual value exchanges}
A~more nuanced discussion requires shifting our perspective
from personal data as solely a~risk or liability,
to personal data as an intentional means
for extracting desired and mutual value
from business-to-customer and other relationships.
We will temporarily reduce subtleties and complexities
to a~neutral interaction model
between an individual and a~service provider,
in order to focus on the basic motives and incentives behind such exchanges.

Whenever an~individual requests a~certain service
from a~third-party provider,
the nature of this service may inherently depend on the availability of
\begin{inlinelist}
  \item relevant data~points
  \item for the purpose of executing the service
        at the requested quality level
  \item during the period of its execution
\end{inlinelist}.
From an efficiency and confidentiality perspective,
we can reasonably expect the individual to prefer
\mbox{\emph{data minimization}}~\cite{Biega2020}---%
in absence of accidental or deliberate barriers
that distort the exposure of additional data
into the more simple or attractive option~\cite{Choi2018,Utz2019}.
Minimization involves limiting the necessary exchange to
\begin{inlinelist}
  \item those data~points strictly required
        for any proper execution of
  \item only the specifically requested service within
  \item solely the time and space required to execute the service
\end{inlinelist}.
In terms of the CIA~triad,
successful minimization thus trades
the \emph{smallest possible decrease in confidentiality}
for the \emph{smallest possible increase in availability}
to complete the requested service,
with \emph{possible compromises on integrity} for convenience.

To illustrate a~strict application of this model,
we follow the example of a~printed magazine subscription.
This service consists of a~transaction
in which the customer pays a~monetary amount
to obtain one year of monthly postal deliveries
of a~specifically chosen magazine series.
We leave the initial payment process out of scope,
and focus on the completion of the delivery process.
Disruption to confidentiality is minimized
when the magazine publisher
only stores the postal address of the customer,
their choice of magazine,
and the end~date of their subscription---%
all of which are securely deleted
after the delivery of the final magazine.
Theoretically speaking,
these few data points suffice
for service providers to offer the subscription service
without additional data collection or processing.
Each party acts rationally and selfishly in this exchange:
the customer prefers receiving the magazine
over keeping their money plus address to themselves,
and vice versa for the provider.

However efficient it may seem,
observe that minimization neither prevents
nor mitigates \emph{all} privacy-adjacent risks.
Given that personal information was exchanged,
one can easily contend the customer's privacy has been compromised,
regardless of our inclination to argue \enquote{necessarily so}.
Although neither having nor breaching privacy
is the key underlying motive for either participant,
the combination of voluntariness with minimization
cannot prevent misuse in even the simplest of cases.
The minimal data~set in this example
still enables the aforementioned burglars
to target subscribers of high-tech luxury magazines
who live in remote~areas.
Unlike a~more limited singular privacy axis,
the broader realm of the CIA triad
encompasses each customer's individual right
to decide whether their continued passion for technology
justifies a~slightly elevated danger of burglary.
Nevertheless,
correctly applied minimization reduces the risk~level
by shrinking the attack surface to the bare minimum
for those who choose to purchase the~service.

In practice,
we realistically know from experience that
the scope of data to which service providers seek access
tends to be larger---sometimes considerably---%
than the strictly minimal scope
required to deliver a~requested service,
implying that other motives are at play.
Financial incentives might drive them to bypass data minimization
with extensions of
\begin{inlinelist}
  \item collected data~points,
  \item purpose, and/or
  \item time and space
\end{inlinelist}.
Perhaps the provider operates a~different business model,
in which their primary source of revenue is not subscriptions
but third-party services based on extended subscriber data.
An extension of purpose occurs when
the provider retargets subscription data
for different objectives,
such as marketing for its other periodicals.
Providers commonly retain customer data
past the end date of their subscriptions,
even if only as a~service to themselves
for advertising or speeding up future subscriptions.
Such nonessential extensions expand the attack surface
or prolong its exposure,
leading to a~heightened risk of various kinds of misuse---%
and the provider's own violation of minimality
already constitutes a~first instance thereof.

From the simple exchange model,
two broad modi~operandi thus emerge:
\begin{enumerate}
  \item Service providers for whom
  the execution of directly requested services
  is the exclusive source of long-term income.
  If we naively presuppose an impeccable inflow of customer data,
  their mission lacks financial incentives
  to process any data beyond the scope of each service's execution.
  This first category counts a~limited number of members today,
  as few companies restrict the storage of customer data
  to the scope of explicitly requested services.

\item Service providers who supplement their long-term gains
  by repurposing data gathered in the context of one service
  for the realization of other services,
  offered to either themselves or different classes of customers.
  Most companies reside in this second category,
  as their business model is predicated
  on their ability to leverage data for additional services.
  Given the high cost of data acquisition at any point in~time,
  they opt for an extended scope in advance,
  even though excessive in the moment.
\end{enumerate}
This chasm could be closed with complementary strategies.
A~punitive approach considers
what barriers and obstacles could be imposed onto the second group
to nudge or force a~substantial portion of~them
into behaving like the first,
which ideally remains unaffected by the intervention.
A~supporting approach instead wonders
which barriers and obstacles could be removed for the first group
in order to encourage voluntary movement by the second,
increasing the former's attractiveness
for all parties involved.

\subsection{Stuttering data incurs a~socioeconomic cost}
Our theoretical examination of intentional data exchange
clearly needs to be augmented
to take ethics and legality into account.
Rather than subjectively imposing a~moralistic value judgment
on different profit~models,
we neutrally observe that
\begin{inlineabc}
  \item an individual may want to request
        a~service from a~provider,
  \item exchange of certain personal data might be a~precondition
        to the requested service's satisfactory fulfillment, and
  \item the provider may independently possess an incentive
        to process data beyond the service's scope of requirements
\end{inlineabc}.
For the sake of argument,
we will hereafter assume
an ethical point of view that considers non-minimal data usage
to cause negative long-term societal effects,
given supporting proof of recent years.
Such conclusion would regard it a~lawmaker's duty
to ensure that
the long-term benefits of data~minimization to service~providers
significantly outweigh
any short- or long-term competitive advantage they would stand to gain
from intentional disregard.
Legislation can achieve this
by decreasing the benefits of undesired behaviors
and/or
increasing the benefits of desired behaviors.

The \emph{General Data Protection Regulation (GDPR)}~\cite{gdpr_2016}
partially was a~reaction to
an exponential spread of unsustainable data~practices,
which it sought to discourage through procedures
that may end~in considerable fines~\cite{wolff_early_2021,sun_gdprxiv_2023}.
Parties involved in personal data processing
must maintain a~capacity for executing certain legal processes,
which may subsequently produce evidence
of either \emph{compliance} or \emph{non-compliance}.
Unfortunately,
such capacity comes at a~non-negligible cost in either case,
leading to a~perverse backfiring in which
companies that can afford the fine of non-compliance
preemptively write it~off as a~business expense.
Through this penalizing interpretation,
the GDPR delivered on its promise of a~level playing field
chiefly in the sense that
it leaves individuals and most companies
now equally powerless vis-à-vis multinationals,
for whom 4\% of annual global revenue
makes a~judicious investment.

Frankly---%
after suppressing the urge to raise a~moralizing finger---%
we must confirm this choice as an astute conclusion
from the GDPR's chosen codification of the ethical connection
between long-term societal cost and legal consequence
as a~short-term economic expense.
Companies act rationally
and have a~duty towards their shareholders
to behave in financially predictable and responsible ways.
When parents incentivize dishwashing
by threatening an 80-cent allowance reduction for untidiness,
an obedient teenager will gladly spend their remaining €19.20
on anything more fun than a~sink filled with dirty~plates.
Coincidentally headquartering the European subsidiaries of many global data~processors,
Ireland grew bizarrely concerned
about being perceived as an overly strict~parent by its dearly beloved adoptees.
It therefore trials a~more reassuring form of compassion
that rewards their offspring's admirable dedication to business~studies~\cite{Lancieri2022}.
Suffice to say that European lawmakers' presumed ethical ambition
of leveraging gentle punishment
to encourage lasting alignment between societal and economic incentives,
turned out to be no~match for cunning lobbyists whose stakeholders
favored a~covert~divergence all~along~\cite{Atikcan2019}.
No need to publicly declare the end of privacy
when you can openly have your cake
and eat someone else's, too.

As a~possibly even more unfortunate side-effect,
the same legislation ultimately hinders benevolent data exchanges
in which all participants sincerely pursue mutual benefit
without hidden data~exploitation.
Whatever motivation remains for being a~data-minimizing company
is usually nipped in the bud
by the inflated cost and risk of
handling \emph{any} amount of data in the first place.
Companies may find avoiding data~minimization more prudent altogether,
because the delta in effort between
minimal and more-than-minimal data processing
tends to~be dwarfed by both options' common cost threshold for data acquisition.
Notwithstanding the GDPR's citizen-empowering aspects
such as \emph{consent withdrawal} and the \emph{right to erasure},
the unbridled continuation of societally detrimental data practices
casts a~long shadow over the EU's worldwide data~stage~\cite{Naughton2018}.
Similar ramifications can be observed
in the more recent \emph{AI~Act}~\cite{ai_act},
whose attempt at balancing
a~commitment to protection
with a~desire for innovation
risks hurting both,
and in several non-European legislative initiatives
such as the proposed \emph{Brazil AI Bill}~\cite{brazil_aibill_2023},
India's \emph{Digital Personal Data Protection Act}~\cite{india_dpdp_2023},
and
Tanzania's \emph{Personal Data Protection~Act}~\cite{tanzania_pdpa_2022}.

In light of these failing protection mechanisms,
one could be tempted to conclude that
personal data nowadays flows all too easily.
However,
we claim the exact opposite.
\emph{Personal data does not flow~well~enough,}
which explains why we are plagued by an abundance of shortcuts,
wherein counting potential fines as an integral part of the data acquisition cost
proves to be the cheaper and hence most rational option.
Rather than an unforeseeable cause,
such calculated behavior is a~symptom of legislative failure,
unfolding within an economic reality
that cannot afford to capture society's thriving data and AI momentum.

The extreme dysfunctionality of our personal data streams
damages society and humanity
immeasurably far beyond diminished business efficacy.
Many obstructions to data propagation
that initially appear praiseworthy through a~superficial privacy lens,
become shallow excuses for denying people a~legitimate choice
of trading a~minimal loss in confidentiality
for an impactful presence of availability.
The consequences of data immobilization
can be life-defining.
Ensuring that data from a~wearable fitness bracelet
reaches a~medical professional on~time,
can mean the difference between life or death~\cite{Lubitz2022}.
Cars unwittingly store evidence about
their owner's quietly progressing Alzheimer's disease,
years before a~general practitioner suspects~it~\cite{Bayat2021}.
Those impacted would violently disagree
that their data ever flows too well.

In~light of such monumental potential,
we cannot help but wonder
how reimagined data flows
could bring crucial improvements to our lives
that outlast the inevitable hurt they cause---%
which starts by dismantling this incessant paradox
as a~false dichotomy long past its expiration date.
Cases such as the ones above
substantiate our perspective that
addressing the profound scarcity of robust data~flows
contributes to obsoleting a~lesser but persistent abundance
of harmful data~flows.
Balanced access to~data
serves both the individual
and those who handle their data
for the delivery of requested services.
Studying mutual benefit
can help us identify sweet~spots
within the CIA~triad.
Nonetheless,
we require an actionable legal approach
to turn such theoretical explorations
into an operational data exchange framework.

\subsection{Trust plays the long game}
\label{subsec:TrustIntro}
In this article,
we introduce
a~techno-legal proposal for \emph{evolvable trust}
as a~sustainable mechanism for data exchange
with well-aligned socioeconomic incentives.
We argue the need for a~more long-term view
on the evolving relationship
between a~data~subject and a~data~controller.
A~relational approach can only succeed
if it recognizes the needs and desires of all sides
as valid and important:
individuals have clear requirements in society beyond confidentiality,
and companies have the reasonable goal of
guarding their profit/investment ratio
to maintain a~competitive economic position.
The wide ethical spectrum in~between
allows us to demarcate areas
where those different demands
are not necessarily at~odds,
so we can identify the roads that lead~us there
and mitigate the obstacles along the path.

On \emph{both} sides of the relationship,
trust is at an all-time low~\cite{waldman_industry_2021}.
The frequently repeated sentiment that
people do not trust companies with their data,
is another component to the ongoing psychological projection
within the privacy realm.
After all,
individuals merely reciprocate
the remarkable lack of trust companies bestow upon them.
The intrusive and upfront cookie dialogs
forge a~thinly veiled disguise
for a~company's severe distrust of
whether a~customer would agree---%
at a~\emph{later} point in time---%
to an anticipated need of~data.
The depth of this distrust
conspires with the nonexistent incentives for minimality
to produce an overly complicated user~experience
that ridicules any residual scraps of
the customer's rightful proclivity for data~minimization.
Thus completes the vicious cycle
of the never-ending self-fulfilling prophecy,
in which companies do not trust
that customers will trust~them,
which prompts them to inform their customers
that they, in fact, do not trust them to~trust~them.

\begin{figure}
  \centering
  \includegraphics[width=.5\linewidth]{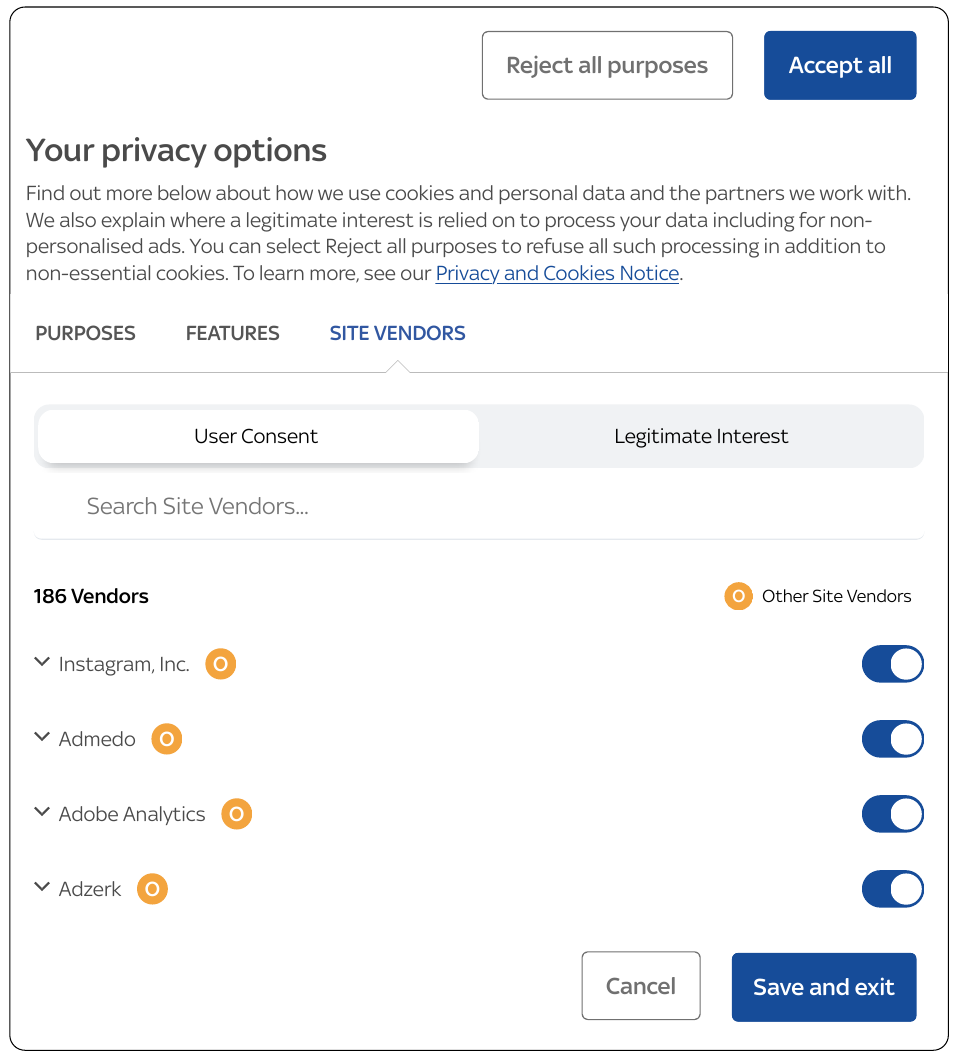}
  \caption{
    Cookie dialogs are up-front agreements of mutual distrust.
    The above specimen (courtesy of Sky~News)
    presents one of the more recent fads where,
    in a~most creative interpretation of the GDPR,
    visitors can seemingly \enquote{consent}
    to the use of legitimate interest
    as a~legal basis for data processing.
  }
  \label{fig:CookieDialog}
\end{figure}

From a~legal perspective,
businesses exploit the unequal power~dynamic
to foist their own substantial compliance burden
upon reluctant consumers,
under the misguided pretense that
what we lack in trust
can unilaterally be substituted by text.
We must frame this stubborn reflex
against a~background of varying scalability properties
of different kinds of one-to-one relationships.
Business-to-business interactions are,
in the grander scheme of things,
magnitudes less numerous
than business-to-customer interactions.
As~such,
businesses more often find themselves in situations
with the time and capability
to tailor legal contracts that can govern their relations.
Advances in machine-assisted contract drafting
further enhance this scalability~\cite{Betts2016}.
In contrast,
the prevailing reliance on \emph{consent}~\cite{custers_role_2022,bourgeus_reinvigorating_2024}
as an unautomatable~\cite{florea_is_2023} legal~basis for data~exchange
encounters scalability hurdles
that increasingly prove insurmountable.
Every website with an unprocessable volume of privacy policies
symbolizes our consolidating despair
via its unique contribution
to the rapidly accumulating costs
for unnamed lawyers and unwilling readers~\cite{McDonald2008}.
They~trigger our daily ritualistic retelling of
\enquote{the biggest lie on the Internet}~\cite{obar_biggest_2020},
wherein we manually pop every last~balloon
that celebrates our society's noble~fight for citizens' rights,
while tacitly wishing we no~longer~had~to.
\cref{fig:CookieDialog} brings an egregious example thereof.

Regardless of people's strongly held moral convictions,
at the end of the day,
the core dynamic of our data-driven society
becomes reduceable to a~straightforward question of economics.
All long-term ethical considerations converge
to constructing an interpretation of a~sustainable balance
between the multiple kinds of monetary and non-monetary costs
paid by participating actors.
The current status~quo holds individuals and companies hostage
in a~self-inflicted Stockholm syndrome,
which has long transgressed all conditions for sustainability,
never~mind~growth.
Let us for once harness this paralyzing chokehold
as a~tremendous opportunity:
any growth impediments
can abruptly escalate into serious business hindrances.
Preparedness to equally consider all parties
in the data exchange relationship
fosters the rare circumstances
under which some might be willing
to escape the local~optimum
that holds us all back.
Could a~focus on mutual value
enable \emph{both} sides of the relationship to~win?

In this article,
we invite readers into a~pragmatic embrace
between the law's noble~intent to embody ethical data~principles
and its simultaneous impotence to prescribe them.
Existing punitive responses to transgressive behaviors
are complemented by our proposal for an unburdening empowerment
that fosters mutually beneficial exchanges
in~a~technology-assisted, scalable~way.
Two-sided cost reduction could spearhead
the realignment of societally advantageous data~minimization
with economically motivated necessity.
We ultimately target the long-term goal of
making responsible data behaviors cheaper
than their currently less expensive unethical shortcuts.

\subsection{Objectives}
\label{sec:Objectives}

Different people across different decades have 
thought about this `trust' conundrum
and reached similar conclusions as we did, 
with the difference that maybe now 
we are in a legal and technical 
landscape that might actually let us do something about it.
Never mind that, in the era of data spaces, 
we also need to piece together the legal jigsaw
that data protection and AI compliance
is becoming, especially in Europe.
This conflictive nature is even present at the most basic level~--~safeguarding
the fundamental right to health can come at the cost of the fundamental right to privacy.
All in all, policy making and technology development
have walked astray for far too long. 
It is long overdue that they shake hands from time to time 
in the hopes that in the near future, they always go hand in hand.

As such, in this article, we 
\begin{inlinelist}
    \item address challenges related to valid GDPR consent,
    \item for scalable relationships between humans, businesses, and machines, and
    \item explore other legal bases,
\end{inlinelist}
towards establishing \emph{long term} evolvable trust
in the exchange of data, services and goods.
Considerations on fundamental facets of trust
are explored in the hopes of establishing
a methodology for the realization of such demands
through an \emph{permanently evolving} relationship.
This approach leverages aspects of 
\begin{inlinelist}
    \item data governance,
    \item automation,
    \item interoperability, and
    \item stakeholder collaboration
\end{inlinelist},
features previously identified as core
for successful RegTech systems~\cite{ryan_gdpr_2021}.
We identify core areas of work 
where these features will be prominent,
namely on
\begin{inlinelist}
    \item compliance and provenance checking,
    \item negotiation on access and usage,
    \item personalized dialogs,
    \item delegation of control,
    \item reusable policies,
    \item auditing, and
    \item secondary data reuse
\end{inlinelist}.

\section{Consent cannot scale}
\label{sec:Consent}





\subsection{The universality of consent}
\label{subsec:universality}

Consent serves as the cornerstone
for processing personal data on the Web~\cite{corren_consent_2022}, 
a principle upheld in both Europe and the United States, 
albeit through contrasting lenses. 
In Europe, the approach is an active one, 
requiring individuals to \emph{opt-in}, 
with the intended goal of giving clear permission 
for their data's use.
Across the Atlantic in the US, the standard shifts, 
embracing \emph{opting-out} as the norm, 
where consent is presumed unless it is objected to.
Despite these divergent practices, 
both regions anchor their legality 
in the fundamental requirement that 
data subjects fully grasp the implications 
of their data's journey and usage.
Furthermore, 
a similar endeavor can be observed globally, 
as new data protection laws spring forth like hotcakes in recent years, 
inspired by GDPR's influence.
This wave of legislative activity 
is often described as the Brussels effect~\cite{bradford_brussels_2019}, 
where the influence of European Union standards 
ripples across the world, 
shaping data privacy frameworks far beyond its borders.

When we consider GDPR's flavour of consent,
its intent to empower the data subject shines brightly.
However, it becomes equally apparent that it falls short
for both individuals and businesses. 
What it brings in empowerment,
it also brings in responsibility,
placing the onus on data subjects 
to stay informed about every aspect of their data's usage. 
This approach brings with it two significant challenges: 
\begin{inlinelist}
  \item a scalability burden, and
  \item an understandability concern
\end{inlinelist},
as legalese can be overwhelming and obscure for individuals.
Looking at its concrete conditions,
consent should 
be provided through a \emph{\enquote{clear affirmative action}},
be \emph{\enquote{freely given, specific, informed and unambiguous}},
\emph{demonstrable by the data controller}, and
\emph{\enquote{be as easy to withdraw as to give}}~(Articles 4.11 and 7~\cite{gdpr_2016}).
These demands also manifest as scalability annoyances for businesses, 
particularly for small and medium enterprises,
which might not have enough resources to
to scale effectively when faced with the challenges of 
meeting auditing duties and 
fulfilling obligations related to consent withdrawal.
Furthermore, when businesses are faced with the decision of selecting a legal ground
for processing the data necessary for their operations,
they inevitably rely on obtaining consent 
to collect as much data as possible during the initial interaction, 
driven by the concern that users may not return in the future.
This underscores, once again, 
that establishing a trust framework 
to facilitate the responsible flow of data 
will yield mutual benefits for both individuals and businesses.
Not only that, the over-reliance on consent mechanisms
also leaves companies at a disadvantage
by not fully realizing the potential of data 
due to consent-imposed constraints.

As such, numerous voices have been raised against the 
efficacy of the consent mechanism to genuinely enhance
the digital lives of people and businesses~\cite{corren_consent_2022}---%
the culture of consent on the Web 
has morphed into a ritual of blind, 
uninformed adhesion to terms of service, 
where decisions are swayed not by understanding 
but by the tides of social pressure 
and collective behavior~\cite{bechmann_noninformed_2014,solove_murky_2023}.
Moreover, beyond placing an impossible-to-deal-with burden on individuals,
consent still is not valid since consent management platforms
and other similar services are not even legally compliant
for them to give informed and specific consent~\cite{noauthor_iab_2024,noauthor_google_2020}.
Furthermore, as~\citeauthor{sun_gdprxiv_2023} highlight,
an examination of GDPR enforcement, 
from its enforcement date to June 26\textsuperscript{th}, 2023,
reveals a substantial body of over 3000 official enforcement documents,
among which 134 documents
reference Article 7 on \enquote{Conditions for Consent}. 
Of those, 65 cases resulted in fines, 
with only 7 imposing penalties exceeding 1 million euros.
This analysis highlights a troubling reality: 
even if not lawful,
the misuse of consent often goes insufficiently challenged 
and it is difficult to challenge effectively.

Venturing beyond the borders of the EU,
and examining the data protection landscape 
in the world's most populous nations, 
the universal reliance on consent becomes strikingly clear.
This reveals an equally pressing need 
to transcend the limitations of consent, 
recognizing that it is but a starting point in safeguarding personal data.
Table~\ref{tab:consent} encapsulates these findings, 
drawing attention to key insights derived from the analysis of
data protection regulations from India (IN), China (CN), US state laws,
Indonesia (ID), Pakistan (PK), Nigeria (NG), Brazil (BR), 
Bangladesh (BD), Russia (RU) and Mexico (MX),
the ten most populated countries in 2023\footnote{
Information derived from \url{https://population.un.org/wpp/} 
(accessed on August 12\textsuperscript{th}, 2024).
}\textsuperscript{,}\footnote{
Country codes and their subdivisions are represented using the Alpha-2 code 
in the ISO 3166 standard, openly available at 
\url{https://www.iso.org/iso-3166-country-codes.html}
}.
Of these,
only Bangladesh stands without a data protection law,  
though its Cybersecurity Act 2023~\cite{bd_cybersecurity_2023}
launched a first step towards legislating the lawful use
of identity information.
Hardly surprising,
given that data,
and especially personal data,
is being treated as a hot commodity these days.
Meanwhile, 
Pakistan's Personal Data Protection Bill (PDPB)~\cite{pk_pdpb_2023}
remains in draft form, yet to solidify into enforceable law.
Following we present a list of the examined laws per country.
\begin{enumerate}
  \item India (IN) -- Digital Personal Data Protection Act 2023 (DPDP Act)~\cite{india_dpdp_2023} 
  \item China (CN) -- Personal Information Protection Law 2021 (PIPL)~\cite{china_pipl_2021} 
  \item United States (US) -- There is no comprehensive national data protection law in the United States, although there are laws focused on specific data types, e.g., consumer health data and minor’s online data, that are excluded from this analysis. Existing state laws, that are already being enforced, are described below.
  \begin{enumerate}
    \item California (CA) -- California Consumer Privacy Act (CCPA)~\cite{usca_ccpa_2018}
    \item California (CA) -- California Privacy Rights Act (CPRA)~\cite{usca_cpra_2020}
    \item Colorado (CO) -- Colorado Privacy Act (CPA)~\cite{us_colorado_2021}
    \item Connecticut (CT) -- Connecticut Data Privacy Act (CTDPA)~\cite{us_connecticut_2022}
    \item Nevada (NV) -- Nevada Privacy of Information Collected on the Internet from Consumers Act (NPICICA)~\cite{us_nevada_2019}
    \item Oregon (OR) -- Oregon Consumer Privacy Act (OCPA)~\cite{us_oregon_2023}
    \item Texas (TX) -- Texas Data Privacy and Security Act (TDPSA)~\cite{us_texas_2023}
    \item Utah (UT) -- Utah Consumer Privacy Act (UCPA)~\cite{us_utah_2022}
    \item Virginia (VA) -- Virginia Consumer Data Protection Act (VCDPA)~\cite{us_virginia_2021}
  \end{enumerate}
  \item Indonesia (ID) -- Personal Data Protection Law 2022 (PDP)~\cite{id_pdp_2022} 
  \item Pakistan (PK) -- Personal Data Protection Bill 2023 draft (PDPB)~\cite{pk_pdpb_2023} 
  \item Nigeria (NG) -- Nigeria Data Protection Act (NDPA)~\cite{ng_ndpa_2023} 
  \item Brazil (BR) -- General Data Protection Law (LGPD)~\cite{br_lgdp_2018}
  \item Russia (RU) -- Federal law No. 152 FZ of 27 July 2006, \enquote{On Personal Data} (the Data Protection Act or DPA)~\cite{ru_dpa_2006} 
  \item Mexico (MX) -- Federal Law for the Protection of Personal Data Held by Private Parties (FLPPDHPP)~\cite{mx_lfpdppp_2010} 
\end{enumerate}

\begin{table}[b!]
  \centering
  \caption[Analysis of consent definitions]{
    Analysis of consent definitions and key requirements in the ten most populated countries in 2023 (the ISO~3166 standard was used to refer to countries and their subdivisions) versus EU's GDPR. Due to the absence of a comprehensive national law, the currently enforceable US state laws were analyzed in its place. Most laws opt for a GDPR-like definition, as showcased by the first nine rows which focus on the definition of consent. The ensuing rows (Opt-out, Notice, Withdrawal, Delegation and Records) highlight key requirements derived from the usage of consent as a legal basis.
    \par
    Since California has two state laws on data protection, fields identified with a X* represent requirements present in both. (X) marks laws where withdrawal of consent is only considered for situations where opt-in consent is required for the processing of sensitive data, as opt-out consent is the norm. Pakistan's law is represented in \textcolor{violet}{X} as it is still a draft and, as such, subject to change. Russia's notice requirements, as identified by X\textsuperscript{DS}, depend on the source of the personal data, if the data subject is the data source, then the notice should be available upon request, while other sources require the notice to be shown before processing. Furthermore, Mexico's data protection authority promotes the availability of different types of notices, X\textsuperscript{3}, depending on the type of data being processed and on how it is notified.
  }
  \label{tab:consent}
  \tabcolsep 2.5pt
  \small
  \begin{tabular}{rcccccccccccccccccc}
    \toprule
      &  &  &  & \multicolumn{8}{c}{\bf US} &  &  &  &  &  &  &  \\
    \cline{5-12}
    \bf Criteria & \bf EU & \bf IN & \bf CN & \bf CA & \bf CO & \bf CT & \bf NV & \bf OR & \bf TX & \bf UT & \bf VA & \bf ID & \bf PK & \bf NG & \bf BR & \bf BD & \bf RU & \bf MX \\
    \midrule
    \bf Affirmative act & X & X &  & X & X & X & X & X & X & X & X &  & \textcolor{violet}{X} & X &  &  &  &  \\
    \bf Freely given & X & X &  & X & X & X &  & X & X &  & X &  & \textcolor{violet}{X} & X & X  &  & X &  \\
    \bf Specific & X & X &  & X & X & X &  & X & X &  & X &  & \textcolor{violet}{X} & X &  &  & X &  \\
    \bf Informed & X & X & X & X & X & X &  & X & X & X & X &  & \textcolor{violet}{X} & X & X &  & X &  \\
    \bf Unambiguous & X & X &  & X & X & X &  & X & X & X & X &  & \textcolor{violet}{X} & X & X &  &  &  \\
    \bf Unconditional &  & X &  &  &  &  &  &  &  &  &  &  &  &  &  &  &  &  \\
    \bf Voluntary &  &  & X &  &  &  & X &  &  & X &  &  &  &  &  &  &  & X \\
    \bf Explicit &  &  & X &  &  &  &  &  &  &  &  &  &  &  &  &  &  &  \\
    \bf Conscious &  &  &  &  &  &  &  &  &  &  &  &  &  &  &  &  & X &  \\
    \midrule
    \bf Opt-out &  &  &  & X* & X & X & X & X & X & X & X &  &  &  &  &  &  & X \\
    \bf Notice & X & X & X & X* & X & X & X & X & X & X & X & \textcolor{black}{X} & \textcolor{violet}{X} & X & X &  & X\textsuperscript{DS} & X\textsuperscript{3} \\
    \bf Withdrawal & X & X & X & (X) & (X) & (X) & (X) & (X) &  &  &  & \textcolor{black}{X} & \textcolor{violet}{X} & X & X &  & X & (X) \\
    \bf Delegation &  & X &  &  &  & X &  & X & X &  &  &  &  &  &  &  &  &  \\
    \bf Records & X & X & X & X* &  &  &  & X &  &  &  & \textcolor{black}{X} & \textcolor{violet}{X} & X & X &  & X &  \\
    \bottomrule
  \end{tabular}
\end{table}

As evidenced in Table~\ref{tab:consent},
it is apparent that the \enquote{Brussels effect}
is more than just a buzzword.
The majority of the analyzed laws 
reflect a definition of consent 
closely aligned with that of the GDPR, 
emphasizing its \emph{freely given}, \emph{specific}, 
\emph{informed} and \emph{unambiguous} nature,
as well as it being an \emph{affirmative act}
taken by the data subject.
Additionally,
if we introduce the notion of consent 
being both \emph{unconditional} and \emph{conscious},
it becomes clear that further reflection is required 
on how such expectations would strain 
already overburdened consent-based systems.

In this context,
of interest is also the choice of US state laws
and their Mexican counterpart,
as they have chosen to adopt \enquote{opt-out} consent as the default approach, 
while reserving \enquote{opt-in} consent as the required legal basis 
for processing sensitive personal data, 
such as health or children's data.
However, it must be stated that
both regulatory approaches grant individuals the right to withdraw consent,
with the exception of Texas, Utah and Virginia.
The same cannot be said about record-keeping.
While \enquote{opt-in} based jurisdictions
require consent to be demonstrable,
their \enquote{opt-out} counterparts mostly do not, explicitly,
identify such a requirement.

The duty to showcase an appropriate consent notice,
enabling individuals to make informed decisions
and allowing businesses to obtain high-quality data,
is prevalent across countries, 
while only Mexico's data protection authority
offers guidelines for three distinct types of notices,
tailored to the nature of the data being processed 
and the method of notice delivery~\cite{mx_lineamientos_2013}, e.g., 
a comprehensive notice is required when data is collected in person,
a simplified version is needed for direct online or phone interactions,
while a short form notice can be delivered when space is limited, 
such as in a SMS.
On that note,
Russia's Data Protection Act~\cite{ru_dpa_2006} notice requirements 
are also distinct from all others:
depending on whether the data is collected from the data subject 
or from other sources,
the duty to provide notice differs~---
in the first case,
an active approach must be taken by the data subject,
as the notice must only be available upon request,
while in the second case,
the law imposes a stricter requirement as
the data subject must be provided with the notice
prior to the beginning of the personal data processing.
This analysis does not delve into the specific contents of the notices, 
which vary across jurisdictions.
While these notices serve as key tools for informing individuals 
when consent is required, 
their function extends beyond consent alone.
Businesses utilize them to meet broader transparency obligations, 
covering various legal grounds for processing data, 
outlining how individuals can exercise their rights, and more. 
Therefore, the intricacies of notice content 
fall outside the primary focus of this section.
It could however be debated whether European policy-makers
could take a page from the other jurisdictions' books 
and allow the usage of distinct types of notices
with more or less information
according to the usage of data being made, its sensitivity,
the preferences of the data subject, and so on and so forth.
Such an approach could benefit both sides of the equation,
sparing users from engaging with details they deliberately choose to avoid, 
while also encouraging businesses to steer clear of the shortcuts 
they currently employ to gain rapid access to data.

Another interesting idea is being put forward in India's DPDP Act~\cite{india_dpdp_2023},
as well as a few of the US state laws~\cite{us_connecticut_2022,us_oregon_2023,us_texas_2023}~---
the idea of having a \enquote{consent manager},
or an \enquote{authorized agent},
acting on the data subject's interest.
While the role of US-based \enquote{authorized agents}
is quite limited in terms of scope,
i.e., they only act on the person's behalf 
to opt-out of processing activities
that do not align with the data subject's preferences,
India's \enquote{consent manager} has a much bigger role to play.
The DPDP Act actually defines that such a manager can act as
\enquote{a single point of contact [...] 
to give, manage, review and withdraw [..] consent through 
an accessible, transparent and interoperable platform}~\cite{india_dpdp_2023}.
Furthermore,
US-based laws put forward techniques to designate such agents
including browser settings or extensions, or
a global setting embedded within an electronic device.
The significance of such techniques cannot be overstated---\emph{every}
business-to-customer interaction inherently
faces greater scalability challenges 
compared to business-to-business interactions
since their are far more numerous and
most customers do not have a proficient capacity
to engage in meaningful negotiations.
Therefore, equipping individuals with robust tools 
that can negotiate on their behalf 
will allow them to manage their personal data 
far more effectively and efficiently.
In Section~\ref{sec:Automation}, we revisit this concept,
highlighting the significant role Web agents will play 
in delegating control over data management 
and enhancing user autonomy.

Regardless of opt-in or opt-out conditions,
or even what information needs to passed in notices
or kept in records,
one fact remains clear~---
consent-as-is as reached its local optimum,
in particular when it comes to empowering users or
making new streams of data available to businesses.
In response to these challenges, 
a growing number of legal scholars have proposed innovative ways 
to get over such a peak.
One prominent suggestion is the concept of broad consent for secondary purposes, 
not specified at the time of consenting, such as
storage or research use~\cite{boers_broad_2015,hallinan_broad_2020}.
While these ideas have been optimistically put in perspective,
they also come with trade-offs,
especially when dealing with stringent requirements surrounding sensitive personal data, 
like health-related data.
The possibility to use
\enquote{Pre-Formulated Declarations of Data Subject Consent}
put forward either by 
data controllers~\cite{clifford_preformulated_2019} or by 
data subjects~\cite{florea_is_2023},
that have agents negotiating or enforcing preferences on their behalf~\cite{slabbinck_enforcing_2024,wright_heres_2024},
is also being discussed as a potential solution,
though both still require advanced protection tools to avoid unfair terms for both parties.


\subsection{Everything, everywhere, all at once}

Over the past decade, 
technological progress has been met with a flurry of policy-making and legislative activity, 
particularly across Europe. 
While the notion of data protection is far from novel~---
its roots tracing back to the 1981's Council of Europe Convention 108+~\cite{council_of_europe_convention_1981} and 
the EU's 1995 Data Protection Directive~\cite{noauthor_directive_1995}~---
the advent of the GDPR has reshaped the global landscape.
This landmark regulation addresses the pervasive use of digital personal data, 
with consent being put at centre stage, 
as underscored in the analysis provided in the previous section.

\begin{figure}[h]
  \centering
  \includegraphics[width=\linewidth]{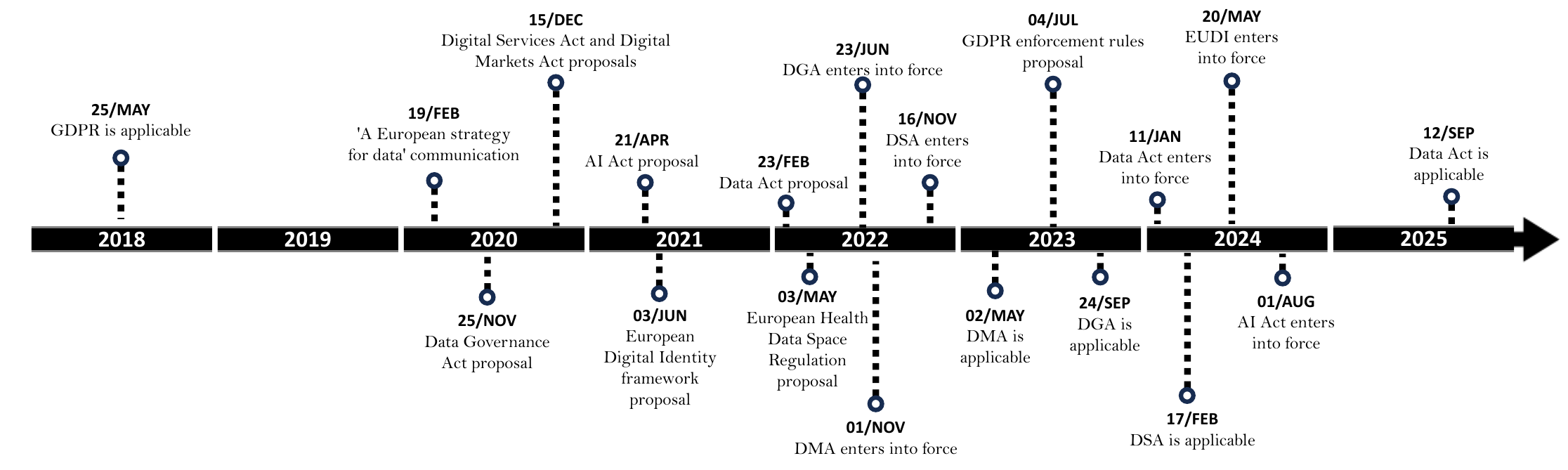}
  \caption{
    Timeline of European data and digital services regulations
    in the awake of the \emph{European strategy for data}.
  }
  \label{fig:laws}
\end{figure}

In the wake of GDPR's global effect, 
inspiring data protection regulations worldwide~\cite{bradford_brussels_2019}, 
the European Commission unveiled its visionary \emph{strategy for Data} in 2020~\cite{european_commission_communication_2020}.
This initiative seeks to weave a seamless flow of data across sectors within the EU, 
while steadfastly upholding European ideals of personal data protection, 
consumer protection legislation, and competition law.
At its core lies a commitment to crafting fair, transparent, and practical rules for accessing and using data, 
all underpinned by mechanisms of trustworthy governance to ensure integrity and balance in the digital age.
A cascade of regulatory proposals has since emerged---many now enacted---each crafted to refine
and expand the landscape of personal (and non-personal) data governance.
From the Data Governance Act (DGA)~\cite{dga} 
to the Digital Services Act (DSA)~\cite{dsa}, 
Digital Markets Act (DMA)~\cite{dma}, 
Data Act~\cite{data_act}, 
European Digital Identity (EUDI)~\cite{eudi}, and 
the European Health Data Space (EHDS) Regulation~\cite{ehds}, 
these measures collectively form a robust framework for the digital age.
Furthermore,
the AI Act~\cite{ai_act} framework also cannot be ignored,
as it regulates the usage of personal and non-personal data to train AI models.
The timeline of these regulations,
from the proposal date to its entry into force and applicability,
is showcased in Figure~\ref{fig:laws}.
With many obligations already in motion---spanning
the DMA's rules for large online platforms, or \enquote{gatekeepers},
the DGA's mandates for data intermediation service providers and data altruism organizations,
and the DSA's framework for digital service providers---the
European strategy for data is advancing with unwavering momentum,
with many already calling for reforms to the DGA~\cite{bobev_white_2023,verstraete_assessing_2023}.
Moreover,
the recently adopted EHDS Regulation cannot be forgotten,
as it imposes restrictions on the use of personal and non-personal electronic health data
whether being for primary care usage or for promoting the secondary use of health data for research, innovation, and policymaking.
Member States will be required to offer at least one digital identity wallet to all citizens and residents by 2026,
as mandated by EUDI~\cite{eudi},
and its Implementing Acts for the core functionalities and the certification of EUDI wallets,
which were recently adopted by the Commission~\cite{eudi_implementing_acts}.
To further underscore the point,
GDPR enforcement rules~\cite{gdpr_enforcement}---additional
requirements to ensure the consistent interpretation and application of the GDPR across the EU---and
GDPR simplification rules~\cite{noauthor_edpbedps_2025}---an
amendment to reduce record-keeping for small and medium-sized enterprises---were
also proposed by the Commission.
Amid the still-unraveled intricacies of these regulations,
and their relationship with the GDPR,
one truth emerges---consent
has been elevated to a central role within this evolving regulatory framework~\cite{chomczyk_penedo_towards_2022}.
As such, in the following paragraphs,
we will explore these regulations and the role that consent plays in them.

\paragraph{European Digital Identity}
The EUDI Framework builds on the electronic Identification, Authentication and Trust Services Regulation (eIDAS)~\cite{eidas}.
The latter establishes a single, interoperable framework for electronic identification and trust services
to enable safer electronic interactions between businesses,
while making them faster and more efficient across the EU.
Furthermore,
trust service providers that comply with eIDAS requirements can also use their services as evidence in legal proceedings.
Examples of trust services are electronic signatures, electronic seals and timestamps, 
website authentication certificates, 
or electronic registered delivery services.
EUDI introduces a series of new requirements in this law,
in particular including the creation of an European Digital Identity Wallet,
which should be certified by certification bodies accredited by data protection supervisory authorities.
Recently, the EC also adopted five implementing regulations related to
\begin{inlinelist}
  \item rules for the integrity and core functionalities of EUDI Wallets~\cite{eudi_core},
  \item rules on the protocols and interfaces of EUDI Wallets solutions~\cite{eudi_interfaces},
  \item rules on person identification data and electronic attestations of attributes of EUDI Wallets~\cite{eudi_attestations},
  \item reference standards, specifications and procedures for a certification framework for EUDI Wallets~\cite{eudi_certification}, and
  \item obligations for notifications to the Commission concerning the EUDI Wallet ecosystem~\cite{eudi_notifications}
\end{inlinelist}.
By default, Wallet providers must remain blind to the details of a user's Wallet-related transactions,
however, an exception arises~--~where
personal data is essential for delivering a Wallet-related service,
the user's explicit consent is required and should be granted for each specific instance.
Additionally,
EU countries are tasked with creating mechanisms at national level
to enable qualified trust service providers in verifying \enquote{qualified electronic attestation of attributes},
i.e., official documents such as, e.g., social security documents or driving licenses.
As such, these providers must be equipped to validate the authenticity of these attributes through authentic sources,
but solely with the informed consent of the individual to whom the attestation pertains.

\paragraph{Data Governance Act}
The DGA~\cite{dga} is one of the key pillars of the \emph{strategy for data}
to create a single European data space.
Its primary aim is to foster the exchange of data across sectors within the EU,
while embedding \enquote{transparency} requirements for data users.
It sets forth conditions for the reuse of specific categories of public sector data and 
introduces a new entity~--~the single information point provider~--~to
facilitate access between data users and this particular data domain.  
Beyond this,
it legislates the market for data intermediation service\footnote{
'data intermediation service' means a service which aims to establish commercial relationships
for the purposes of data sharing between an undetermined number of data subjects and data holders on the one hand
and data users on the other, through technical, legal or other means [...]~(Article 2.11~\cite{dga})}
providers and data altruism\footnote{
'data altruism' means the voluntary sharing of data on the basis of the consent of data subjects
to process personal data pertaining to them [...] without seeking or receiving a reward
that goes beyond compensation related to the costs that they incur where they make their data available
for objectives of general interest [...],
such as healthcare, combating climate change, improving mobility,
facilitating the development, production and dissemination of official statistics,
improving the provision of public services, public policy making or 
scientific research purposes in the general interest~(Article 2.16~\cite{dga})}
organizations,
paving the way for data subjects and data holders to share their information with these entities,
in the latter's case for the \enquote{public good}.
Regarding consent, 
none of these new services can be overlooked.
When contemplating the re-use of data pertaining to data subjects,
public sector bodies must enable the reuse of data on the basis of consent, 
ensuring that data flows only with the deliberate agreement of its subjects.
To this end, they are entrusted with guiding potential re-users in seeking such consent,
providing pathways through technical mechanisms to channel these requests for reuse,
as well as with informing data subjects about the possibility to refuse consent~(Recital 15~\cite{dga}).
Furthermore, data intermediation service providers,
when offering their services to data subjects,
must provide avenues for individuals to wield their rights,
including \enquote{giving and withdrawing their consent to data processing}~(Recital 30~\cite{dga}).
Additionally,
data intermediation service providers that operate as 
data cooperatives\footnote{%
\enquote{services of data cooperatives} means data intermediation services offered by an organisational structure 
constituted by data subjects, one-person undertakings or SMEs [...]~(Article 2.15~\cite{dga})}
aim to \enquote{strengthen the position of individuals in making informed choices before consenting to data use, 
influencing the terms and conditions of data user organisations 
attached to data use in a manner that gives better choices to the individual members of the group or 
potentially finding solutions to conflicting positions of individual members of a group 
on how data can be used where such data relates to several data subjects within that group}~(Recital~31~\cite{dga}).
As for data altruism activities,
to cultivate trust in the process of granting and withdrawing consent with clarity and ease,
a European data altruism consent form will be developed by the European Commission,
with the support of the European Data Innovation Board and 
the European Data Protection Board.
Especially in the realms of scientific research and statistical use of data,
this form should serve as a bridge between legal terms and
a user-friendly experience for altruistic data sharing~(Recital 52~\cite{dga}).
Nonetheless, the latest practical guide related to the DGA,
dropped by the Commission on September 24\textsuperscript{th}, 2024,
does not provide any updates on when the form will be available~\cite{european_commission_implementing_2024}, and
flags are already being raised regarding the economic viability of these newly fledged services~\cite{verstraete_assessing_2023}.
There is, however,
a push for the use of standardized semantic specifications to design these forms,
while maintaining the potential for personalization and contextualization~\cite{rossi_one_2025,pandit_implementing_2024}. 
Currently,
there are at most twenty four data intermediation services
providers\footnote{\url{https://digital-strategy.ec.europa.eu/en/policies/data-intermediary-services}, accessed on August 4\textsuperscript{th}, 2025.}
and two data altruism
organizations\footnote{\url{https://digital-strategy.ec.europa.eu/en/policies/data-altruism-organisations}, accessed on August 4\textsuperscript{th}, 2025.}
registered by the Commission,
while the DGA has been applicable since September 24\textsuperscript{th}, 2023,
and data intermediation services providers need to comply with their obligations by September 24\textsuperscript{th}, 2025.
When it comes to data altruism,
there are ambiguities in its definition and application~\cite{baloup_white_2021}---it focuses on consent, 
although consent is only one of many applicable legal grounds for the processing of personal data,
and the altruistic reuse of non-personal data seems to be implicitly overlooked.
Not only that,
the interpretation of broad consent or how to assess compatible re-uses of data still lacks clarity,
while the notion of data altruism activity does not provide enough detailed information,
in particular in relation to the allocation of responsibilities and
the protection of individuals when it comes to their data rights.
Furthermore,
the Commission has abandoned the idea of having a rulebook for data altruism\footnote{
  \url{https://ec.europa.eu/info/law/better-regulation/have-your-say/initiatives/14094-Data-altruism-Rulebook-for-organisations-collecting-personal-data-in-the-general-interest_en}, accessed on August 4\textsuperscript{th}, 2025.
},
which was set out on DGA's Article 22~\cite{dga} and was supposed
to provide additional information related to the obligations set out for these organisations.
As for data intermediation services, 
several issues have also been pointed out.
Besides the description of potentially demanding conditions for providing such services,
described in DGA's Chapter III~\cite{dga},
the scope of activities and business models of intermediaries beseeches clarification~\cite{baloup_white_2021,verstraete_assessing_2023}.
The interplay of DGA's rules with other branches of law,
e.g., cybersecurity, or price regulations,
must be internally unbundled by organizations seeking to supply intermediation activities,
with clear guidance from the Commission and its supra-national authorities still lacking~\cite{bobev_white_2023}.


\paragraph{Data Act}
The Data Act~\cite{data_act} constitutes a cross-sectoral legislative framework and 
a central pillar of the European data strategy, 
designed to establish harmonised principles governing equitable access to and responsible use of data.
It covers both personal and non-personal data,
while laying down requirements on  
\begin{inlinelist}
  \item making data generated through connected products and related services accessible to the user,
  \item making data from data holders available to data recipients and
  \item to public sector bodies and authorities when needed for public interest tasks,
  \item ensuring fair terms for data sharing contracts between data holders and small and medium-sized companies or not-for-profit research organisations,
  \item enabling the switch between data processing services,
  \item establishing safeguards to protect data holders against unlawful data transfer outside the EU, and 
  \item developing interoperability standards for data to be accessed, transferred and used
\end{inlinelist}.
Moreover,
it reviews certain aspects of the 1996's directive~\enquote{on the legal protection of databases}~\cite{database_directive_1996},
to make data generated or acquired from Internet of Things~(IoT) devices accessible for access and reuse.
While the Data Act is primarily concerned with non-personal data,
in circumstances where users of connected products and related services qualify as data subjects,
the rights conferred by the Act operate in addition to those established under the GDPR~\cite{gdpr_2016}.
In fact,
a central challenge for the Act lies in its interaction with the EU's personal data protection framework,
particularly in contexts where organizations process datasets comprising both personal and non-personal data,
as this distinction is paramount to assess the scope of application of the GDPR~\cite{finck_they_2020}.
Additionally,
the rights of access and data portability set forth in GDPR's Articles 15 and 20~\cite{gdpr_2016}
are further complemented by this Act by allowing users to
\enquote{access all data generated by the use of a connected product, 
  whether personal or non-personal}~\cite{data_act}.
Accordingly,
when wishing to process personal data,
a GDPR legal ground such as consent needs to be in place.
The influence of consent within the framework of Data Act requirements extends far beyond this point---even
when data sharing agreements are automatically executed,
there must be tooling in place to interrupt and terminate them 
with the mutual consent of the involved parties~(Recital 104~\cite{data_act}).
Furthermore,
ensuring coherence with the requirements of the DGA~\cite{dga}, DMA~\cite{dma}, and the AI Act~\cite{ai_act},
as well as with adjacent branches of law---including 
contractual obligations, competition law, and the protection of trade secrets---presents
further potential sources of friction.

\paragraph{Artificial Intelligence Act}
The AI Act~\cite{ai_act} establishes a risk-based approach
to guide the governance of artificial intelligence across the EU, 
assigning responsibilities to providers and users
according to the weight of the risks their AI systems may carry. 
At its sternest,
the Act draws a clear boundary against high-risk applications that endanger human safety: 
systems designed to manipulate,
to score individuals socially,
to remotely identify them, 
or to sway the will of voters are categorically forbidden.
Yet, the Act is not meant not stifle innovation---it
allows general-purpose AI to flourish,
provided that it safeguards the fundamental rights of EU citizens and 
upholds transparency through practices such as AI impact assessments.
Exemptions for research endeavors and 
for components shared under open-source licenses are also granted.
Regarding the role of consent,
the Act extends its scope beyond its usage to justify personal data processing:
\emph{\enquote{informed consent}} from subjects is also required
for the testing of an AI system, and its intended purpose, in real-world conditions,
and following the GDPR counterpart,
it may be revoked at any time without providing any justification.

\paragraph{European Health Data Space Regulation}
The EHDS Regulation~\cite{ehds} is the first legislative initiative in the EU
aimed at establishing a common European data space.
Its objective is to empower individuals by granting them improved access to their electronic personal health data,
both at the national and European levels.
In parallel,
the Regulation seeks to establish governance mechanisms for primary and secondary use of electronic health data, and 
foster the development of a unified market for electronic health record (EHR) systems, 
wellness applications, and medical devices.
In addition to extending the rights of access, rectification, and data portability
established in Articles 15, 16, and 20 of the GDPR~\cite{gdpr_2016},
this Regulation grants data subjects,
specifically with regard to personal electronic health data,
the following rights:
\begin{inlinelist}
  \item to insert information in their own EHR,
  \item to restrict access,
  \item to obtain information on accessing data, 
  \item to opt out in primary use, and
  \item to lodge a complaint with a digital health authority
\end{inlinelist}.
Furthermore, by 26 March 2027,
the EC will launch, through implementing acts,
technical specifications for the priority categories of personal electronic health data, namely
\begin{inlinelist}
  \item patient summaries,
  \item electronic prescriptions,
  \item electronic dispensations,
  \item medical imaging studies and related imaging reports,
  \item medical test results, and
  \item discharge reports
\end{inlinelist}.
These specifications will define the European electronic health record exchange format,
which must be commonly adopted in the EU, machine-readable, and 
enable the transmission of personal electronic health data across different software applications, devices, and healthcare providers.
By that same deadline,
the Commission will also give shape to a central interoperability platform for digital health.
This infrastructure will provide services to enable the exchange of personal electronic health data
among the Member States' national contact points for digital health.
Moreover,
the EHDS Regulation provides a set of measures to ensure a consistent, and trustworthy secondary use of health data.
It establishes the minimum categories of electronic health data
that data holders must make available for secondary use,
such as electronic health data from EHRs or data from wellness applications,
as well as the purposes for which it can be processed,
e,g, improvement of the delivery of care, scientific research related to health, and policymaking and regulatory activities.
Nonetheless,
data subjects have the right to opt out of secondary use at any time without providing any justification.
Be it for primary care or for the secondary use of data,
every activity including the processing electronic health data must rest upon a GDPR lawful ground~(Recitals 15 and 52~\cite{ehds}).
As such, within this framework,
consent emerges once more as a pivotal subject matter,
particularly in light of the sensitivity that health data embodies.



\paragraph{}
The misuse and overuse of consent,
as a~generic mechanism to exchange personal data on the Web,
is a well-known fact by now.
Not only that,
the transactional speed at which it is happening
also serves as a telling that it cannot be considered valid consent.
With the key role that consent was given within the EU's data and technology regulatory landscape,
one might wonder whether the issues that
consent has been facing as a legal basis for personal data processing
are not being seriously debated.
It seems that we are already past what consent can make possible
for both individuals and companies,
and now we are asking even more to fulfil
the well-desired common European data spaces.
In this context,
we will explore the advantages and disadvantages of going beyond consent,
exploring the myriad of other lawful grounds that allow the processing of personal data.

\section{Beyond consent}
\label{sec:BeyondConsent}

\subsection{The lawful six}
\label{sec:LawfulSix}

Consent is only but one of the lawful grounds prescribed by the GDPR
for data controllers to process personal data of EU citizens.
Beyond the oft-cited ground of consent,
Article 6~\cite{gdpr_2016} provides a framework of alternative legal bases,
each carrying its own rationale and applicability.
\emph{Contractual necessity} enables data processing when it is necessary
for fulfilling contractual agreements in which the data subject is a party.
\emph{Legal obligation}, by contrast, reflects the authority of law itself,
compelling controllers to comply with its provisions.
\emph{Legitimate interest} occupies a more nuanced position,
inviting a balancing of organizational aims against the rights and freedoms of individuals.
The ground of \emph{public interest} allows processing where the welfare of society outweighs private considerations,
including the activities of official authorities.
Finally, \emph{vital interest}, more rarely invoked,
stands as a safeguard in moments of urgency for the protection of life or safety of data subjects.

The determination of an appropriate legal basis for processing under the GDPR
is inherently dependent on the specific purpose of the processing activity,
as well as the nature of the data involved and other contextual factors
such as the relationship between the controller and the data subject.
The choice of legal ground must therefore be carefully aligned with the underlying rationale for processing and
the risks associated with the data at issue.
Importantly, when these conditions evolve---for example,
if the purposes of processing change---it becomes necessary to re-evaluate the chosen legal basis and
determine whether a new one must be established to ensure continued compliance.
Table~\ref{tab:ComparisonGrounds} presents a comparative analysis of the previously identified lawful bases,
evaluating them across several dimensions of practical relevance.
Specifically,
it considers their potential for automation, that is,
the extent to which a given legal basis may enable organizations to automate or semi-automate access to personal data.
It further examines the degree of involvement required from the data subject,
highlighting variations in the level of individual intervention.
Finally, the table assesses scalability,
understood as the ease with which a legal basis can be consistently applied at scale across multiple services and jurisdictions.
These dimensions are particularly relevant for organizations seeking to streamline compliance processes
while ensuring that individual rights remain adequately protected.
Furthermore,
this analysis applies exclusively to categories of personal data that
do not fall within the GDPR's definition of special category data,
for which additional safeguards and legal conditions are required.

\begin{table}[ht]
  \centering
  \caption{Comparison of GDPR legal grounds in terms of automation, data subject interaction, and scalability.}
  \begin{tabular}{rlll}
    \toprule
    \bf Legal basis         & \bf Automation potential & \bf Data subject interaction & \bf Scalability \\
    \midrule
    \bf Consent             & Low                      & High                         & Low             \\
    \bf Contract            & High                     & Medium-Low                   & High            \\
    \bf Legal obligation    & High                     & Low                          & High            \\
    \bf Legitimate interest & Semi                     & Medium                       & Medium          \\
    \bf Public interest     & Low                      & Low                          & Low             \\
    \bf Vital interest      & Semi                     & None                         & Low             \\
    \bottomrule
  \end{tabular}
  \label{tab:ComparisonGrounds}
\end{table}

From the perspective of \emph{automation potential},
contract and legal obligation emerge as the most amenable to deterministic, rule-based implementations.
Contractual processing can often be operationalized in a straightforward manner,
as access decisions follow directly from the predefined terms of the agreement between the parties.
Similarly, legal obligation lends itself to automation where obligations are clearly codified,
though complexity may arise when overlapping or conflicting legal regimes demand context-specific interpretation.
Legitimate interest and vital interest, by contrast, allow only partial automation:
both require human oversight, either through a contextual balancing test in the case of legitimate interest,
or through predefined emergency scenarios in the case of vital interest.
Consent and public interest remain the least conducive to automation.
Consent would require highly granular data subject preferences
to account for the diverse range of purposes, data types, and data controllers involved,
while public interest processing is typically confined to specific authorities and
thus not easily generalized across contexts.

The dimension of \emph{data subject interaction} further illustrates the varying operational implications of each legal basis.
Consent requires the highest level of engagement,
encompassing opt-in mechanisms,
ongoing policy management to manage one's preferences,
and the possibility of withdrawal.
Legitimate interest involves a more limited form of interaction,
namely the right of the data subject to object to processing.
Contractual processing entails interaction at the point of sign-up or acceptance,
after which further data subject involvement is minimal.
Legal obligation and public interest grounds typically preclude any opt-out,
as compliance with statutory or societal requirements takes precedence over individual choice.
Vital interest, by definition, applies in circumstances where the data subject is unable to act,
thereby eliminating direct interaction altogether.

In terms of \emph{scalability},
contract and legal obligation again stand out.
Standardized contractual terms make it possible to extend the same legal ground across large numbers of users,
while legal obligations enable the construction of compliance pipelines around well-defined statutory requirements.
Legitimate interest is somewhat more limited,
but still more scalable than consent,
since it avoids the administrative overhead of user-by-user consent collection.
While the balancing test introduces interpretive complexity,
standardized assessment tools could partially mitigate this burden.
By contrast, consent suffers from inherent scalability challenges,
mostly related to its informed and specific character.
Public interest and vital interest grounds are even more narrowly circumscribed:
the former applies predominantly to specific public bodies,
while the latter is invoked only in exceptional, often life-threatening situations.

In conclusion,
the analysis suggests that the contractual legal ground offers significant potential for both automation and scalability,
as its reliance on predefined agreements allows organizations 
to implement standardized agreements as the basis for rule-based personal data processing activities,
while contract law does not forbid such arrangements,
given that the legal limits when processing special data are strictly respected.
Contract is also a recognized and operative legal basis in several EU laws beyond just the GDPR,
namely under the Data Act~\cite{data_act} to use data generated by connected products.
Similarly,
the legal obligation ground demonstrates strong suitability for automated and scalable processing,
though its applicability remains confined to contexts where
regulatory compliance explicitly requires the handling of personal data.
Hence,
both lawful grounds should be used where consent should not due to inherent power imbalance between controller and subject,
for instance when employers are processing the personal data of their employees~\cite{article_29_data_protection_working_party_guidelines_2018}.
Legitimate interest, by contrast,
with its broad and not purpose-specific scope~\cite{article_29_data_protection_working_party_opinion_2014},
should be considered to enable exchanges that can create value for both controllers and data subjects,
even if they override the latter's interests at times.
Accordingly,
the performance of balancing exercises in every distinct context is important to avoid misuse~\cite{morel_legitimate_2023}.
Overall,
broadening the range of lawful bases beyond consent
enables data controllers to maintain compliance with the GDPR,
while simultaneously alleviating individuals from the considerable scalability and comprehensibility challenges
that consent would pose if it were required for all personal data processing activities encountered in daily life.

\subsection{Jurisdictional flavors}
\label{sec:JurisdictionalFlavors}

While the EU's GDPR establishes a comprehensive framework with six primary legal bases,
other regions either adopt similar structures or deviate by introducing unique grounds or
limiting their scope of applicability.
The comparative overview in Table~\ref{tab:World} illustrates both the convergence and
divergence of legal bases for personal data processing across major jurisdictions---previously
described in Section~\ref{subsec:universality}.

\begin{table}[ht]
  \centering
  \caption{Legal basis from the several data protection regulations described in Section~\ref{subsec:universality}, including the EU's GDPR for comparison. X is used to identify the legal bases present in each regulatory text. O is used to differentiate opt-out consent from its opt-in counterpart---all US state laws only include opt-out consent. In the case of India, (X) marks legal bases that are covered by the \enquote{Legitimate uses} legal basis present in the DPDP Act~\cite{india_dpdp_2023}. Pakistan's law is represented in \textcolor{violet}{X} as it is still a draft and, as such, subject to change.}
  \begin{tabular}{rcccccccccccc}
    \toprule
    \bf Legal bases & \bf EU & \bf IN & \bf CN & \bf US & \bf ID & \bf PK & \bf NG & \bf BR & \bf BD & \bf RU & \bf MX \\
    \midrule
    \bf Consent & X & X & X & O & X & \textcolor{violet}{X} & X & X &  & X & O \\
    \bf Contract & X &  & X &  & X & \textcolor{violet}{X} & X & X &  & X &  \\
    \bf Legal obligation & X & (X) & X &  & X & \textcolor{violet}{X} & X & X &  & X \\
    \bf Vital interest & X & (X) & X &  & X & \textcolor{violet}{X} & X & X &  & X &  \\
    \bf Public interest & X & (X) & X &  & X &  & X &  &  &  &  \\
    \bf Legitimate interest & X &  &  &  & X & \textcolor{violet}{X} & X & X &  & X &  \\
    \bf Legitimate uses &  & X &  &  &  &  &  &  &  &  &  \\
    \bf Legally disclosed &  &  & X &  &  &  &  &  &  & X &  \\
    \bf Disclosed by DS &  &  & X &  &  &  &  &  &  & X &  \\
    \bf Required by law &  &  & X &  &  & \textcolor{violet}{X} &  &  &  & X &  \\
    \bf Public administration &  &  &  &  &  &  &  & X &  &  &  \\
    \bf Research &  &  &  &  &  &  &  & X &  & X &  \\
    \bf Rights exercise &  &  &  &  &  &  &  & X &  &  &  \\
    \bf Credit protection &  &  &  &  &  &  &  & X &  &  &  \\
    \bf Media and journalism &  &  &  &  &  &  &  &  &  & X &  \\
    \bottomrule
  \end{tabular}
  \label{tab:World}
\end{table}

As previously established, in most jurisdictions,
\emph{consent} serves as a foundational basis,
whereas in the US state laws and Mexico,
it operates primarily as an opt-out mechanism with certain opt-in requirements for sensitive data.
Contract and legal obligation also exhibit substantial alignment across several jurisdictions.
In the EU, Brazil, Russia, Indonesia, Pakistan, Nigeria, and China,
\emph{contracts} provide a legitimate basis where the processing is necessary for contractual performance.
\emph{Legal obligation}, similarly,
is a consistent ground across these frameworks,
except for China, 
though its interpretation may differ.
Beyond legal obligation with court orders or statutory duties,
in China, Pakistan and Russia,
personal data can be processed if \emph{required by other laws or administrative regulations}.

Certain jurisdictions expand the list of recognized grounds with context-specific provisions absent from the GDPR.
India's DPDP Act~\cite{india_dpdp_2023} introduces \emph{legitimate uses},
encompassing situations such as voluntary data provision, medical emergencies, public health,
employment, or legal obligations,
leaving out contractual necessity or legitimate interests.
China explicitly recognizes personal data \emph{disclosed by the individual} or made \emph{legally public},
with the Russia's DPA law~\cite{ru_dpa_2006} providing akin bases.
Brazil introduces additional categories including \emph{research},
\emph{exercise of rights} in judicial, administrative or arbitration proceedings,
\emph{credit protection},
and \emph{processing by public administration} for policy execution.
Russia similarly codifies provisions for \emph{media and journalism}, as well as research.
These specific, context-driven legal bases reflect a pragmatic accommodation of
socio-economic and institutional needs within national data protection regimes.

At the same time,
\emph{public interest} and \emph{vital interest} remain unevenly adopted.
The EU, China, Indonesia, and Nigeria acknowledge these categories,
particularly for tasks carried out by public authorities or in emergency situations,
but their absence in other jurisdictions is either folded into more specific categories,
e.g., legitimate uses in India or research in Brazil and Russia,
or addressed through sector-specific rules rather than general data protection law.

A striking divergence emerges in the United States and Mexico,
where opting-out is the norm.
Opt-in is reserved for the handling of sensitive data, including child data,
in the California, Colorado, Connecticut, Oregon, Texas, and Virginia state laws,
while Nevada's opt-in consent mechanism concerns only health data,
and the Mexican law goes one step further by mandating written consent for all sensitive personal data.
This structure underscores a consumer protection orientation rather than a legal basis framework in the European sense.

Taken together, these findings reveal a fragmented but partially convergent global landscape.
While consent, contract, and legal obligation form common denominators across many systems,
several jurisdictions supplement these with additional, context-specific grounds.
Such variations reflect differing legal traditions, regulatory priorities, and socio-political contexts,
ultimately shaping how data controllers and subjects experience the balance between flexibility, compliance, and protection.
Nonetheless,
the contractual legal basis warrants particular emphasis,
as it not only enables greater automation and scalability compared to consent,
but also provides a lawful foundation for processing personal data across a wide range of jurisdictions,
thereby offering long-term benefits for both controllers and data subjects.
In the following section, we delve into the concept of \emph{evolvable trust}---a
scalable techno-legal alternative that explores GDPR's legal bases
without overreliance on consent as the sole lawful ground,
while simultaneously establishing economic incentives
for the exchange of data, goods, and services among the parties involved.

\section{Trust as a~temporal relationship}
\label{sec:Relationship}
\subsection{Defining trust}
Despite abundant mentions of the word \enquote{trust},
we have so far not defined it,
as such definitions tend to stray into philosophy.
To mitigate this risk,
we present an operational definition
for the purposes of this article:

\emph{\textbf{Trust} is a~relationship between individual people
  or groups of individuals,
  the quality of which evolves over time,
  serving as a~channel that supports the exchange of
  data, services, and/or goods.}

By defining trust as an interpersonal relationship,
we explicitly exclude relationships between people and devices or software;
the idea being that people fundamentally cannot bestow trust upon technology,
for technology can impossibly experience or reciprocate trust.
As such,
in our model,
people trust companies,
even if such companies are represented by software.
The role of technology is not that of a~party in the trust relationship,
but rather to support such relationships at scale.

Current technologies are not well equipped
to support the inevitable evolution
that characterizes the dynamics of human trust relationships.
As such,
the potential for improvements on each side of the relationship
is substantial.
We compare the current, limited model
with a~model that better reflects the complexity of the real~world.

\subsection{Brief Explosion Of Trust}
\label{subsec:beot}

Within theoretical physics,
the Big Bang model postulates a~single dense moment
at the beginning of time,
which produced all matter that can ever exist
within this universe.
A~careful analysis of current practical GDPR interpretations
reveals that most practitioners collectively decided
the exchange of data must necessarily adhere
to a~similar one-shot model of~time.

\begin{figure}
  \centering
  \includegraphics[width=.9\linewidth]{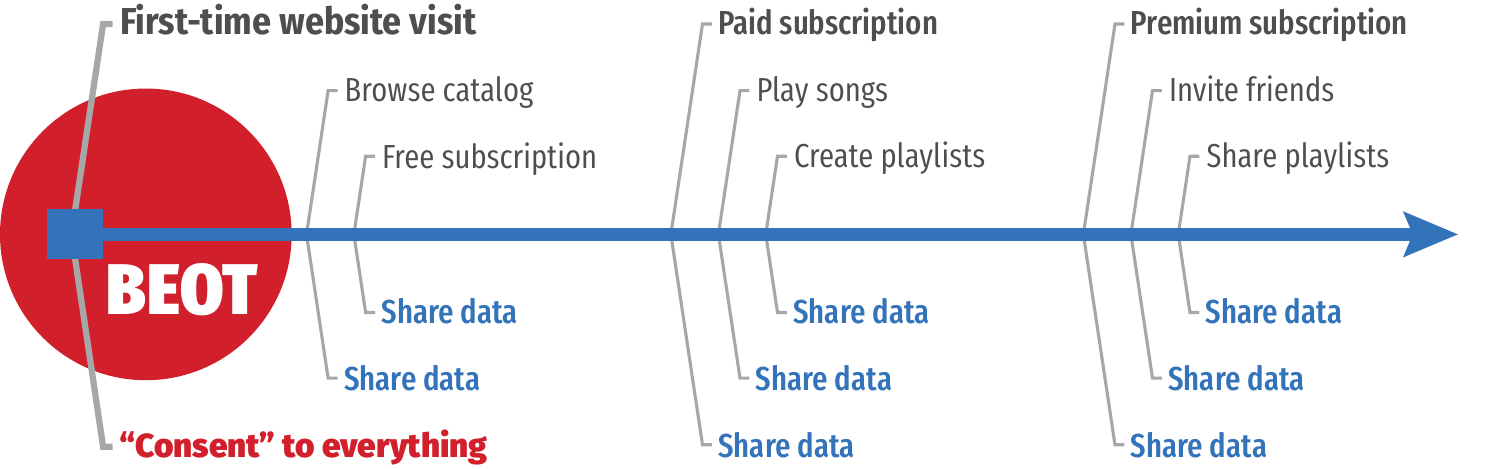}
  \caption{
    The dominant timeline for digital trust relationships
    starts with the \emph{Brief Explosion Of Trust (BEOT)},
    at which point a~data~subject
    is asked to agree to an exhausting list of demands,
    in stark contrast to the absence of a~factual trust relationship
    that would normally serve as collateral.
    Informed decision or negotiation is impossible
    as the relationship itself has not yet commenced,
    leaving no opportunities to assess trustworthiness.
    All needed and unneeded conditions are established beforehand,
    as the consent dialog model disregards
    the evolutive character of real-world relationships.
  }
  \label{fig:TimelineBeot}
\end{figure}

The majority of data controllers therefore adopt the \emph{BEOT}~model
featuring a~\emph{Brief Explosion Of Trust}
in which they consider
a data subject's first moment of interaction
as the immediate and final opportunity
to obtain any possible guarantees and permissions
across all of spacetime.
\cref{fig:TimelineBeot} depicts a~timeline within this model,
through an example in which a~person builds a~relationship
with a~provider of streaming music.
Note how the initial visit of the website
corresponds to the point where the data subject
has to make all decisions about trust for the entire relationship,
despite not having started said relationship.
The clear advantage is that
every imaginable future act of sharing data
\emph{will} be covered by the single prior agreement;
the obvious disadvantage that it \emph{has} to be,
resulting in an agreement that is overly and unnecessarily wide
at every single point on the timeline where data is shared.

Despite its roots in physics,
the BEOT~model demonstrates an existential allergy to Schrödinger's cat
through its intolerance for uncertainty
as to whether a~given data point is available or unavailable.
The underlying angst stems from the presumption that
anything left uncreated
at the start of the joint timeline with a~data subject
might never come into existence.
Assuming this model of the data universe,
a~controller's only rational behavior consists of
seizing the initial moment of the~BEOT
to demand a~data subject's unconditional and non-negotiable consent
for all possible data usages,
thereby already anticipating any possible timeline developments
for themselves or their current and future partners.
As a~result,
the data subject is asked to a~settle all of the trust
that could be possibly be required for the entire duration of the relationship
before the relationship takes place---%
by essentially agreeing that trust is indeed absent.

To simplify their detection,
each BEOT is announced by an increasingly longer consent dialog
describing the extent of the explosion
at a~level of detail beyond the understanding of most.
Data subjects consequently live through BEOT experiences
in the Old English sense of \emph{bēot}:
a~ritualized promise to proclaim one's acceptance
of a~seemingly impossible challenge
in order to gain tremendous glory for its accomplishment.

Sadly,
the accomplishment itself limits both parties,
as the ability to evolve is crucial to B2C~relationships.
Let us therefore not forget that a~lack of trust
incurs a~significant cost for people and companies.
From a~legal angle,
it is highly doubtful whether the threshold
of informed consent is achieved.
Nonetheless, the agreement texts themselves
likely provide some degree of protection
due to their sheer length,
which could make litigation prohibitively expensive.
At the same time, they paint a~grim picture
of the times to come,
especially for businesses who aim to establish a~genuine relationship
for mutual benefit---%
which not coincidentally forms the cornerstone of economics.
Companies cannot leverage trust as a~key differentiator
with a~static interaction model
that imposes fixed boundaries at the start.

\subsection{Evolvable trust}
\begin{figure}
  \centering
  \includegraphics[width=.9\linewidth]{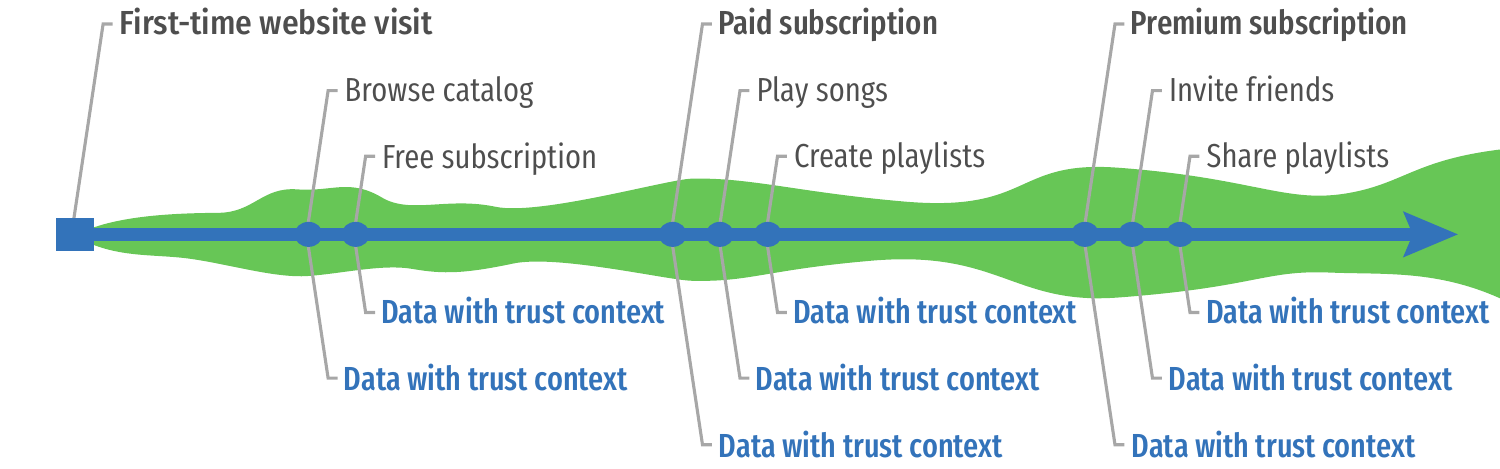}
  \caption{
    In an evolvable trust model,
    no decision or exchange happens
    at the first moment of interaction;
    rather, the relationship is allowed and encourage to grow.
    Every act of exchanging data is encapsulated in a~trust context
    pertaining to this specific data point and exchange,
    reflecting the status of the relationship at that moment in~time.
    This model aligns with the dynamics of real-world relationships,
    notably due do its elimination of widely scoped prior agreements
    regarding abstract future interactions
    that cannot be meaningfully decided yet.
  }
  \label{fig:TimelineTrust}
\end{figure}
In a~model of evolvable trust,
the qualities of trust in the model
need to reflect those of the real-world
at any given point in time.
Specifically, the following conditions need to hold:
\begin{itemize}
  \item There is no presumption of trust
        at the start of the relationship.
  \item The quality of trust can evolve
        in a~positive or negative direction
        at any point in time.
  \item Every act of exchange is contextualized
        with specific policies
        that explicitize the understanding of trust
        at the moment.
\end{itemize}
\vskip.5\baselineskip

As an illustration of those principles,
the timeline in \cref{fig:TimelineTrust}
reimagines the interaction of \cref{fig:TimelineBeot}
in which a~person establishes a~new relationship with a~streaming music provider.
The relationship starts when the person visits the website,
at which point there exists no notion of trust between both parties.
Through interaction with the website,
the person gains confidence that the company can provide value to~them,
yet no exchange has taken place yet.

By browsing specific parts of the music catalog,
the person discloses the first piece of personal data,
namely their interest in certain artists and genres.
This specific act of exchange in isolation
could be argued from the perspective of legitimate interest,
since disclosing a~preference for a~certain genre
is a~necessary precondition to display it.
As such,
the data point \enquote{this person prefers 80s rock music}
is exchanged with the context that
the person considers usage of this data
for the selection of songs during the active browsing session
to be legitimate interest.
At this point,
the person can back out of the relationship
with the understanding that their exchange of data
will not impact them any further,
given the specific scoping in time and purpose.
The company has achieved its goal of demonstrating
they provide a~service relevant to the person's musical tastes.

The company subsequently offers a~free subscription,
which can be activated by creating an account
through a~combination of email address and password.
These individual data~points constitute personal information,
which can similarly be exchanged
with a~specifically scoped legal basis.

This mechanism is invoked repeatedly
so support the evolution of the relationship at every step.
When a~stronger engagement is required,
such as when the person decides to purchase a~paid subscription,
explicit consent will be asked
for the credit~card information to be used
for this specific transaction.
At every point in the relationship,
each exchanged data point is accompanied
by the minimal trust context required
to achieve every step.
The advantage of this approach is that
the company never requires a~level of trust or agreement
beyond what is strictly necessary,
and that all the information to make an informed decision
is available to the person at the point where they have to decide.
The disadvantage is,
compared to the single action pattern of the BEOT,
the large number of individual decision points
during the course of the relationship timeline.
However,
their vastly increased volume is compensated
by a~significant narrowing in scope for each point,
opening~up the possibility of assistance or even automation.

\subsection{Assisted or automated trust decisions}
\label{sec:Automation}

The development of assisted and automated trust systems requires a rethinking of
how individuals, organizations, and machines
interact in the exchange of data, services, and goods.
Evolvable trust cannot be sustainably operationalized
under the current paradigm of data usage management if,
instead of a BEOT,
millions of decisions have to be taken by companies and people alike.
To move beyond this impasse,
a higher level of abstraction is necessary
for the representation of data usage conditions.
Rather than requiring individuals to engage with the minutiae of every data transaction,
meta-policy languages and policy catalogs can provide scalable mechanisms
for defining, deriving, and instantiating policies that align with
individual preferences, legal requirements, and organizational needs.
These higher-order policy frameworks would allow for contextual instantiation
at the moment of decision-making,
enabling the selection of appropriate legal bases---including
but not limited to consent---under
the GDPR and other regulatory systems.
Based on these policies,
both assistance and automation mechanisms could support individuals and companies
in the evolution of their relationship.

Such mechanisms require dynamic negotiation protocols
if they are to support the exchange of a data point with a trust context
that includes \emph{specific} data usage conditions.
Negotiation protocols,
expressed in standardized and machine-readable formats,
enable individual preferences and controller requirements to be reconciled efficiently,
independently of the legal ground that justifies such a transaction.
To ensure that both individuals and companies
can meaningfully engage with these negotiations,
personalized dialogs are indispensable. 
Under consent,
this means tailoring information to support informed choices,
whereas in the context of legitimate interest,
dialogs may highlight the right to object.
Personalized interfaces,
when adaptive and responsive to individuals' distinct wishes for control,
enhance autonomy and ensure that reliance on any given legal basis is both valid and transparent.
The delegation of control further extends this approach
by enabling individuals to entrust decision-making to other individuals or automated agents.
For example,
consent managers or intelligent policy agents can enforce predefined user preferences
while adapting to evolving contexts.
India's \emph{consent manager}~\cite{india_dpdp_2023} and
US-based \emph{authorized agents}~\cite{us_connecticut_2022,us_oregon_2023,us_texas_2023}
represent regulatory efforts that incentivize the use of such technologies.
Complementing these measures,
auditing mechanisms ensure that assisted and automated processes remain accountable over time.
Integrated auditing functions,
strengthen both ex-ante and ex-post accountability,
reassuring individuals that their data is handled as promised
while equipping organizations with demonstrable evidence of compliance.

Taken together,
these areas illustrate that assisted and automated trust decisions
must operate across multiple layers of governance, communication, and oversight.
Their integration provides not only a~scalable and interoperable infrastructure for data exchange
but also a pathway to establishing evolvable trust relationships that endure over time.



\section{Conclusions}
\label{sec:Conclusions}
Above all,
consent dialogs represent an astonishing lack of imagination
between technologists, lawyers, and economists.
On the one hand,
we deem technology so powerful
that profit maximization equates unlimited data extraction
by all legal and adjacent means available.
On the other hand,
we perceive the same technologies
as fundamentally limited in their capability
to capture the complex relationship dynamics
that supposedly necessitate the processing of such vast amounts of data
in the first place.
The presumed inevitability of explosive moments of distrust
reaffirms the lack of competitive advantage
rather than addressing it responsibly.

Trust is not a~zero-sum game.
Hence,
it should not be understood as a~struggle over who exercises control,
but as a~right for self-determination within relationships
to the benefit of all involved.
Genuine autonomy emerges when individuals can continuously influence
how their data contributes to an evolving relationship,
without being forced to choose between abstract notions of absolute privacy and complete exposure.
In contrast to a~one-time binary consent model,
trust functions as a personalized balance between competing values
such as
confidentiality and availability,
conciseness and transparency,
accountability and privacy,
all of which can be adjusted according to each relationship over time.
Responsible personalization dissolves the impasse of lopsided benefits
by tailoring control to how a~specific person
can advance their own interests as well,
rather than imposing uniform rules that work for neither individuals nor companies.

Implementing evolvable trust relationships requires systems
that reduce the cost and effort of acting in the ethically preferred way.
People must be able to delegate nuanced choices to tools
that act in their interest,
where both human and machine actors share a clear understanding
of contextual appropriateness.
These systems model trust not merely as a~layer atop consent,
but as the broader container within which permission and consent evolve.
It all starts with the notion that \emph{both} parties in a~relationship
stand to gain \emph{more} from a~proportional exchange,
in a~striking parallel to the core economic principles of trade.

Yet as we sacrifice the red~herring of absolute privacy
on the altar of untapped socioeconomic potential,
let us remember---%
akin to the biblical miracle of the multiplying bread and fish,
or Disney's \emph{Mickey~Mouse}
after the 2023 copyright expiration of \emph{Steamboat Willie}---%
that data above all remains a~copyable good,
and exhibits vastly different properties
compared to other basic elements
needed to comfortably live our lives.
Unlikely Rumpelstiltskin,
whose life depends on the secrecy of a~single data point,
we are continuously empowered to leverage our own data
thanks to this inherent regenerative capacity.
Increasing the scalability of the legal mechanisms
must be predicated upon the notion that encouraging better data~flows---%
as opposed to solely punishing questionably shortcuts---%
can realize proportional benefit the individual named therein.
As much as widespread mistreatment of data
has caused unfathomable damage to people and society,
individuals' life-changing capability
currently locked within their own data
is still infinitely bigger and more valuable
than what any company can squeeze out of there.
With a~significant amount of potential gain present in every data point;
progress does not equal halting the extraction of all gains,
but rather the reappropriation of these gains
to those the data personally affects.
Our economic incentives therefore require urgent realignment
with our socio-ethical considerations of long-term sustainability,
which ultimately involves appropriate codification into~law.
May an increased understanding of
trade-offs between confidentiality and availability,
together with the vast automation opportunities
for beyond-consent data exchanges,
lead us to more responsible data~flows for mutual benefit,
so~help~us~GDPR.

\printbibliography

\end{document}